\documentclass[prb,a4paper,twocolumn,floatfix,nopacs,showkeys,amsmath,amssymb,nobibnotes,altaffilletter]{revtex4}

\usepackage{graphicx}
\usepackage{pslatex}
\usepackage{xspace}
\usepackage{subfigure}

\newcommand{\degree}{$^{\circ}$}

\begin{document}

\preprint{}

\title{Polymeric Squaraine Dyes as Electron Donors in Bulk Heterojunction Solar Cells}

\author{Sebastian F. V{\"o}lker}
\author{Shinobu Uemura}
\affiliation{Institut f{\"u}r Organische Chemie, Universit{\"a}t W{\"u}rzburg, Am Hubland, D-97074 W{\"u}rzburg (Germany)}

\author{Moritz Limpinsel}%
\author{Markus Mingebach}%
\author{Carsten Deibel}%
\affiliation{Lehrstuhl f{\"u}r Experimentalphysik VI, Universit{\"a}t W{\"u}rzburg, Am Hubland, D-97074 W{\"u}rzburg (Germany)}

\author{Vladimir Dyakonov}%
\affiliation{Lehrstuhl f{\"u}r Experimentalphysik VI, Universit{\"a}t W{\"u}rzburg, Am Hubland, D-97074 W{\"u}rzburg (Germany)}
\affiliation{Functional Materials for Energy Technology, Bavarian Centre for Applied Energy Research (ZAE Bayern), D-97074 W{\"u}rzburg (Germany)}
\affiliation{Wilhelm-Conrad-R{\"o}ntgen Research Center for Complex Materials Systems (RCCM), W{\"u}rzburg}

\author{Christoph Lambert}\email{lambert@chemie.uni-wuerzburg.de}%
\affiliation{Institut f{\"u}r Organische Chemie, Universit{\"a}t W{\"u}rzburg, Am Hubland, D-97074 W{\"u}rzburg (Germany)}
\affiliation{Wilhelm-Conrad-R{\"o}ntgen Research Center for Complex Materials Systems (RCCM), W{\"u}rzburg}

\date{\today}

\begin{abstract}
A polysquaraine low band gap polymer was synthesized by Yamamoto coupling of a monomeric dibromo indolenine squaraine dye. The resulting polymer has a weight average molar mass in the order of $M_w\sim$30.000-50.000 and a polydispersity of ca. 1.7 as determined by gel-permeation chromatography (GPC). The electronic properties of monomer and polymer were investigated by cyclic voltammetry, absorption and emission spectroscopy. Owing to exciton coupling the absorption bands of the polymer are red-shifted and strongly broadened compared to the monomer squaraine dye. Bulk heterojunction solar cells were prepared from blends of the polysquaraine with the fullerene derivative [6,6]-phenyl C61-butyric acid methyl ester (PCBM) in different weight ratios (1:3 to 1:1). The power conversion efficiencies under simulated AM 1.5 conditions yielded 0.45 \% for these non-optimized systems. The external quantum efficiency (EQE) shows that the photoresponse spans the range from 300 to 850 nm, which illustrates the promising properties of this novel organic semiconductor as a low band gap donor material in organic photovoltaics. 
\end{abstract}


\keywords{conjugated polymers, dyes / pigments, polysquaraines, synthesis, UV-vis spectroscopy}

\maketitle

\section{Introduction}

Organic bulk heterojunction solar cells are based on two active components (electron donor/electron acceptor) that are mixed together in order to form a thin film with a microscopic phase separation. As electron donors, small molecule dyes or conjugated polymers are used. As electron acceptors, small molecule dyes, conjugated polymers, fullerene or its derivative PCBM ([6,6]-phenyl C61-butyric acid methyl ester) are commonly used.[1-4] Bulk heterojunction solar cells that are based on conjugated polymers such as poly-3-hexyl-thiophene (P3HT) and electron acceptors such as PCBM achieve remarkable power conversion efficiencies of up to $\eta$ = 6 \% for highly optimized systems with an external quantum efficiency (EQE) exceeding 60 \% within their absorption range of 300 nm to 700 nm.[5,6] However, they are still inferior to inorganic solar cells based on crystalline silicon, where efficiencies beyond $\eta$ = 20 \% are attained.[7] The key concept of these bulk heterojunction solar cells is the formation of a microscopic phase separation of donor and acceptor materials which leads to a large surface contact area at the heterojunction. There are several critical factors such as the open-circuit voltage, $V_{oc}$ (which depends on the HOMO(donor)-LUMO(acceptor) difference), the short circuit current, $J_{sc} = ne\mu E$ (which depends on the density of charge carriers $n$ with elementary charge $e$ and their mobility $\mu$ in the electric field $E$) and the fill factor, $FF$ (which is the maximal attainable power $V_{max}$ $J_{max}$ of the solar cell devided by a given $V_{oc} J_{sc}$ product; $FF$ depends on the lifetime of charge carriers and the series resistance of the cell). The maximal attainable efficiency $\eta=V_{max} J_{max} /P_{light}$ is determined by the spectral distribution of the incident photon-to-current efficiency, $EQE = (1240 J_{sc})/(\lambda P_{light})$. The widely studied solar cells with poly[2-methoxy-5-(2'-ethyl-hexyloxy)-1,4-phenylene vinylene (MEH-PPV) or P3HT in combination with PCBM as the acceptor suffer from the lack of absorbance in the red and NIR region.[8,9] Because of this deficiency, there is an ongoing search for better polymers that absorb down to ca. 900 nm. In this work, we present a promising new low band gap polymer that is based on an indolenine squaraine dye which absorbs light from 300 nm to 900 nm and already reaches 0.45 \% power conversion efficiency in non-optimized systems with PCBM. 

Squaraines represent a class of organic dyes that gets more and more into the spotlight. They can be found in numerous applications and are used as photoconductors, as reporter molecules for SERRS spectroscopy (Surface Enhanced Resonance Raman Scattering Spectroscopy), for data imaging and as nonlinear optical substances.[10-14] Because of their unique absorption and fluorescence properties they play an important role as highly stable fluorescence labels and as detectors for cyanides, mercury ions or sulfide ions.[15-22] Furthermore, there are reports about polysquaraines with low band gaps of less than 1.0 eV.[23-25] First papers that describe squaraines for the use in photovoltaic devices date back to the 70ies.[26,27] Since then, several groups have published reports with squaraines applied in dye sensitized photovoltaic cells.[28-38] In 2008, Silvestri et al. reported the first organic bulk heterojunction cell running on a low molecular weight squaraine/PCBM system with promising results and efficiencies up to 1.24 \%.[39] Since then, a few more publications revealed increasing efficiencies of bulk heterojunction cells with single molecule squaraine/PCBM systems.[40,41] Until now, all the photovoltaic devices based on squaraines use molecular squaraine units. However, polysquaraines have not been explored successfully for the use in photovoltaic devices yet. Since molecules that are based on a cyanine-like donor-acceptor-donor-system (which is the case for squaraines) typically have a sharp and intense absorption in the red to near-IR region, squaraines should have best prerequisites to fulfil the requirements for the use in a photovoltaic device as far as the band gap is concerned. As Grahn et al.[42,43] have shown some years ago, oligomerization of squaraines leads to even smaller band gaps which would allow harvesting of low-energy sunlight.[36] However, since the $S_2$ state in squaraines is forbidden by symmetry, there is a decent gap of absorption at higher energies.[44]

In the present work we describe the polymerization of a squaraine dye by Yamamoto coupling and investigated the potential of the polysquaraine as a donor-component in organic bulk heterojunction solar cells. We will show that owing to exciton coupling, the band gap of the polysquaraine is further decreased compared to the monomer squaraine and the absorption band is strongly broadened which partially fills the absorption gap mentioned above.

\section{Results and Discussion}

\subsection{Polymer Synthesis and Characterization}

\begin{figure*}
	\includegraphics[width=12.0cm]{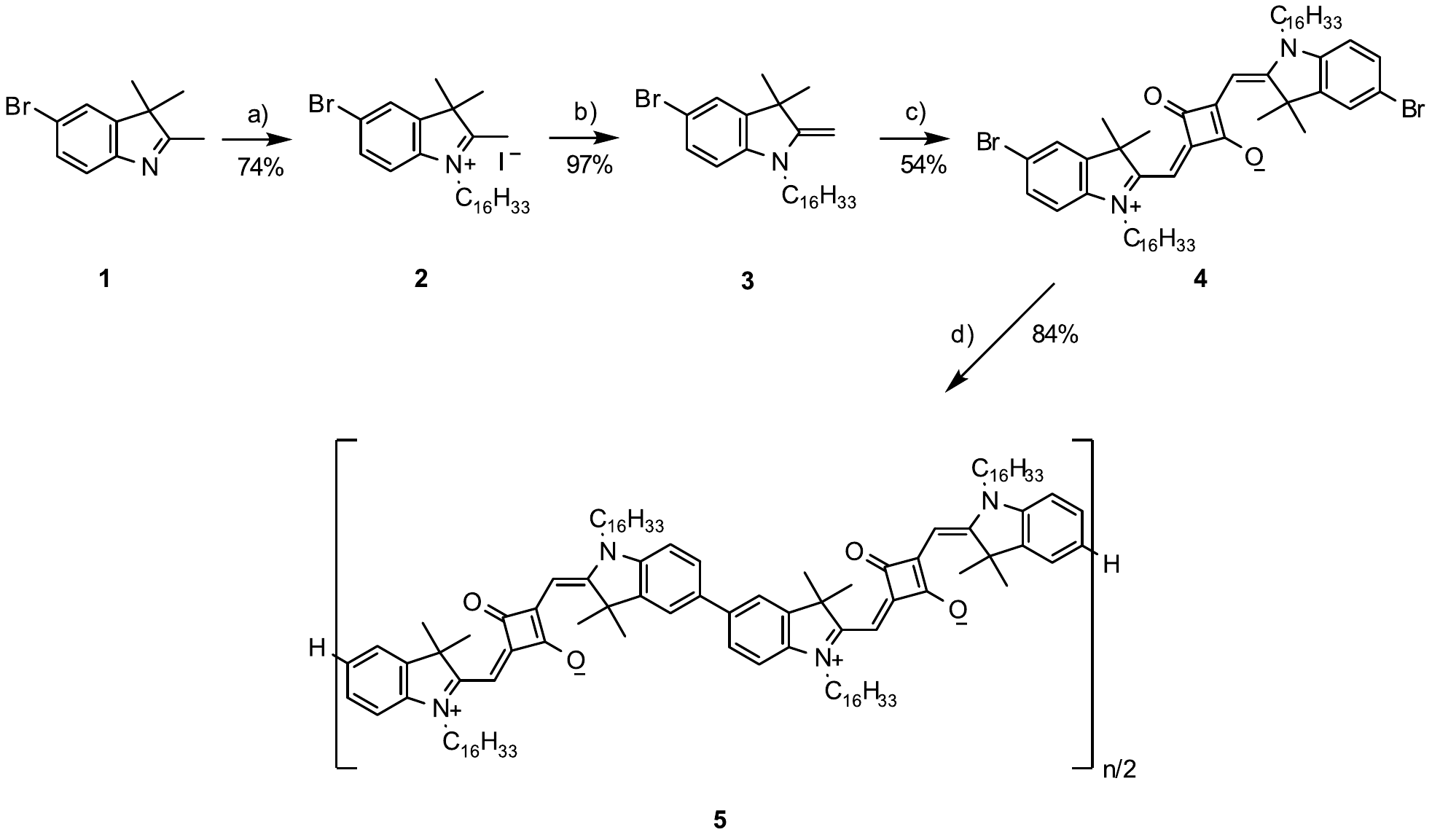}%
	\caption{a) 1-Iodohexadecane, MeNO2, reflux, 15 h; b) 2N NaOH; c) 0.5 equivalents of squaric acid, toluene/n-butanol 1:1, reflux; d) Ni(1,5-cyclooctadiene)2, 1,5-cyclooctadiene, 2,2?-bipyridine, DMF/toluene, rt, 6d.%
	\label{scheme:1}}
\end{figure*}

While polysquaraines usually are prepared by condensation of a bis-donor-functionalised aromatic compound with squaric acid, we used monomeric dibromosquaraine which will directly be polymerized via aromatic C-C-coupling. Suitable symmetric monomeric dibromosquaraines were prepared by Grahn et al. before, and we followed their synthetic route which proved useful.[42,36] We started with 5-bromo-2,3,3-trimethyl-3$H$-indole {\bf 1}, which was synthesized according to known literature procedures.[45] $N$-Alkylation of the 3$H$-indole was carried out with 1-iodohexadecane in nitromethane solution. Since low solubility is a well known issue for such polymers, we chose a long alkyl chain moiety for the $N$-alkylation. Deprotonation of the resulting quaternary salt {\bf 2} with NaOH gave the corresponding methylene base {\bf 3}. The squaraine {\bf 4} was prepared by condensation of squaric acid and two equivalents of the methylene base in a 1:1 mixture of toluene and $n$-butanol using a Dean-Stark trap to remove water. 

Polymerization of the dibromosquaraine {\bf 4} was carried out at room temperature applying the Ni-catalyzed Yamamoto coupling in a solvent mixture of dimethyl formamide (DMF) and toluene. The reaction was terminated by the addition of MeOH/HCl. This resulted in terminal H-atoms. The resulting absence of bromine was confirmed by mass spectrometry. Several consecutive Soxhlet extractions were performed with increasing polarity of the solvents to remove left over reagents and shorter polymer chains. The polymer ({\bf 5}) was obtained as a purple solid after six days in good yields (84~\%).

The solubility of the polymer was rather limited. Various solvents were tested but good solubility in a pure solvent was only achieved in chloroform and low solubility was achieved in dichloromethane. Furthermore, the polymer was soluble in solvent mixtures of various alcohols such as methanol, ethanol, butanol with either dichloromethane, 1,3-dichlorobenzene or 1,2,4-trichlorobenzene.

Mass spectrometry (MALDI-TOF) of the polymer revealed a series of signals with nearly equal distances at around [n $\times$ 843.7]+2 [m/z] up to a molecular weight which corresponds to $n$ = 12 monomeric units. The average mass difference of 843.7 [m/z] corresponds to a single monomer unit without bromine atoms, 2 [m/z] corresponds to the terminal H-atoms. There is no sign of partially brominated polymers at all.

Gel permeation chromatography (GPC) in chloroform with polystyrene as standard was performed to define the molecular mass distribution. It revealed a polydispersity of D = 1.72, a number average molar mass of $\bar{M}_n$ = 28000 and a weight average molar mass of $\bar{M}_w$ = 48100 which corresponds to $\bar{X}_w\sim$57 units. Since it is known that the GPC reveals higher mass values for rod like polymers, we used a conversion factor of 1.42 which was empirically attained by Bunz et al. when comparing the molecular weight of a poly(dialkyl-p-phenyleneethynylene) with its reduced poly(2,5-dialkyl-p-xylylene) by GPC.[46] Using this factor we obtained a corrected number average molar mass of $\bar{M}_n$ = 19700 and a weight average molar mass of $\bar{M}_w$ = 33900 which corresponds to $\bar{X}_w\sim$ 40 units. This was necessary not only to get a more accurate mass distribution but to calculate the extinction coefficient of the polymer more precisely. Furthermore, the narrow $^1$H-NMR signals and the lack of signals for the terminal protons are in agreement with a long polymer chain.

\subsection{Absorption and Emission Spectroscopy}

\begin{figure}
	\includegraphics[width=8.0cm]{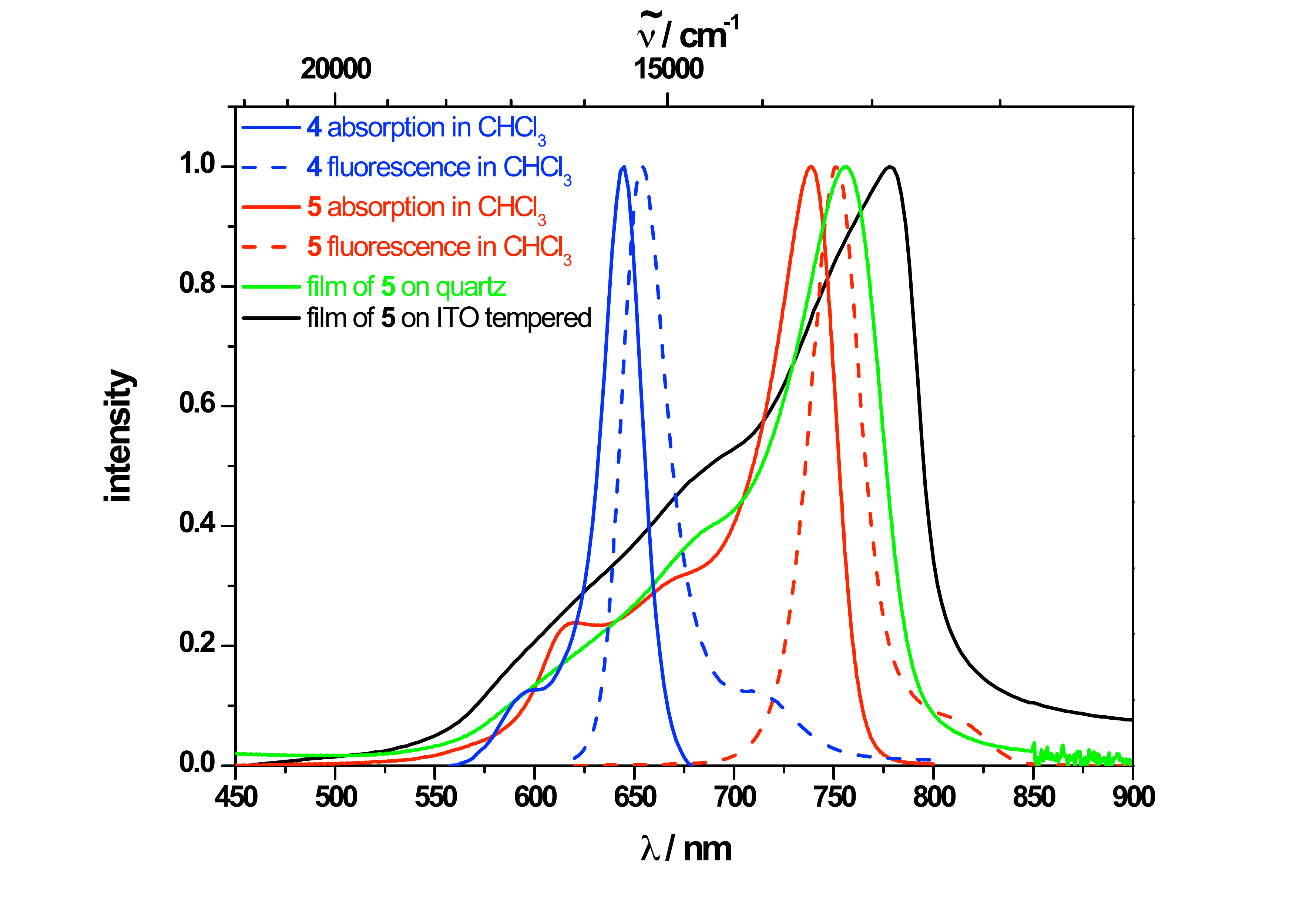}%
	\caption{Normalised absorption and emission spectra of monomer {\bf 4} and polymer {\bf 5} in chloroform and absorption of {\bf 5} as a film on a quartz substrate and of {\bf 5} as a film on an ITO substrate after tempering at 130\degree{}C for 10 min.%
	\label{fig:1}}
\end{figure}

Absorption and fluorescence measurements were carried out for the monomeric squaraine {\bf 4} and the polymer {\bf 5}. Furthermore, the transition moments were calculated according to eq.~\ref{eqn:1}[47]:
\begin{equation}
	\mu_\text{eq}^2 = \frac{3hc\epsilon_0 \ln 10}{2000\pi^2 N}\frac{9n}{(n^2+2)^2} \int \frac{\epsilon}{\tilde{\nu}} d\tilde{\nu}
	\label{eqn:1}
\end{equation}

The monomer shows an intense absorption of $\epsilon = 3.62 \times 10^5$ l mol$^{-1}$ cm$^{-1}$ in CHCl$_3$ ($\mu$ = 11.6 D) in the visible region at 646 nm (0,0 band) (Figure~\ref{fig:1}) with vibronic shoulders at the higher energy side typical of cyanine type dyes. This transition is polarized along the long molecular axis. Upon polymerization the absorption maximum shifts by 2000 cm$^{-1}$ to 738 nm ($\epsilon = 1.90 \times 10^5$ l mol$^{-1}$ cm$^{-1}$ in CHCl$_3$ per monomer unit) and the signal broadens considerably. The transition moment $\mu$ of the whole low energy band is 12.2 D per monomer unit. 

\begin{figure}
	\includegraphics[width=8.0cm]{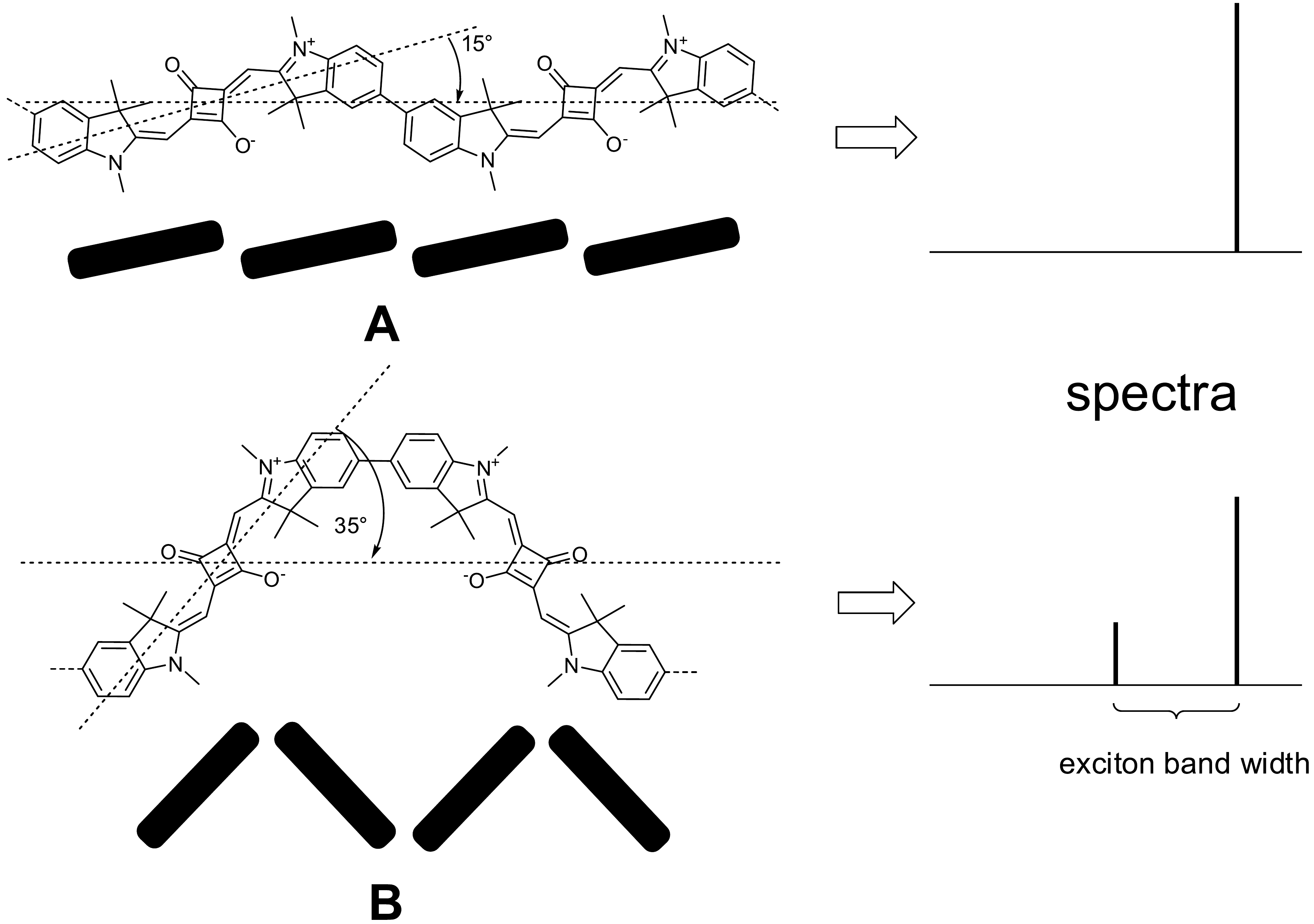}%
	\caption{Idealized polymer sections with elongated herring bone structure ({\bf A}) and zig-zag structure ({\bf B}) and expected absorption spectra.%
	\label{fig:2}}
\end{figure}

In analogy to the interpretation of the shifts and broadening of absorption bands in thiophene-bridged squaraine oligomers we interpret the absorption features in {\bf 5} being due to exciton coupling.[43] Two different possible idealized polymer structures are conceivable (see Figure~\ref{fig:2}): a linear chain in which the monomers (and thus the polarization of the allowed lowest energy transition) form a herring bone structure with an angle ? relative to the vector connecting the centers of the monomers (structure {\bf A}), and, alternatively, a zig-zag chain ({\bf B}) with angle $\alpha$. For the former, exciton coupling theory predicts a single allowed transition into the lowest energy exciton state (J aggregate) with intensity $N1/2\mu$, where $N$ is the number of monomer units and $\mu$ is the transition moment of the monomer. The exciton band width $\Delta E$ (difference of lowest and highest exciton energy) is given by eq.~\ref{eqn:2} in the point-dipole approximation where $r$ is the center to center distance of the monomer in the polymer.[48] 

\begin{equation}
	\Delta E = E_\text{highest} - E_\text{lowest} = 4 \left| \frac{N-1}{N} \cdot \frac{\mu^2}{r^3} \left(  1 - 3 \cos^2 \alpha \right) \right|
	\label{eqn:2}
\end{equation}

For the alternative zig-zag arrangement, both excitations into the highest and the lowest exciton band are allowed, their intensities are $\mu_\text{lowest}= N^{1/2} \mu \cos \alpha$ and $\mu_\text{highest}= N^{1/2} \mu \sin \alpha$, respectively, and the bandwidth is given by eq.~\ref{eqn:3}.

\begin{equation}
	\Delta E = E_\text{highest} - E_\text{lowest} = 4 \left| \frac{N-1}{N} \cdot \frac{\mu^2}{r^3} \left(  1 +  \cos^2 \alpha \right) \right|
	\label{eqn:3}
\end{equation}

From Figure~\ref{fig:1} it is obvious that there is a transition (621 nm) at even higher energy than the monomer 0,0 peak. Thus, polymer model {\bf A} can be excluded as the only conformer present in solution. It is more likely that a mixture of different conformers with both sections of type {\bf A} and sections of type {\bf B} are present. The latter would explain the observation of the peak at 621 nm which then refers to the highest exciton band as expected for a {\bf B} type polymer. For both polymer types, the bandwidth corresponds to four times the electronic exciton coupling energy $V$ if $N\rightarrow\infty$. Assuming an angle $\alpha$ of ca.\ 15\degree for {\bf A} and of 35\degree for {\bf B} (from AM1 optimizations) gives roughly the same exciton band width. Thus, irrespective of the actual presence of either type {\bf A} or {\bf B} polymer section we can derive the electronic coupling from the energy difference of the highest and lowest transition peak which is 2560 cm$^{-1}$ and, thus, leads to $V = \Delta E/4 = 640$ cm$^{-1}$ as an average for the polymer {\bf 5}. The point dipole approximation (eq.~\ref{eqn:3} with $r = 15$~\AA and $\mu = 11.9$ D) yields $V = 370$~cm$^{-1}$ for type {\bf B} conformer and $V = 330$~cm$^{-1}$ (with eq.~\ref{eqn:2} and $r = 16$~\AA) for type {\bf A}. Although these values are about half of that derived from the observed bandwidth they essentially support the excitonic model. However, we cannot exclude that direct conjugational effects also add to the observed absorption bandwidth in the polymer.

The fluorescence spectra of the compounds {\bf 4} and {\bf 5} (Figure~\ref{fig:1}) were measured in chloroform solution. The monomeric squaraine shows an emission spectrum typical of cyanine like dyes, that is, a sharp emission with maximum at 654 nm being symmetrical to the absorption spectrum with its maximum at 646 nm and with a small Stokes shift (200 cm$^{-1}$). The fluorescence quantum yield $\Phi_{fl}$ is 0.49. 

The emission spectrum of polymer {\bf 5} does not represent a mirror image of the absorption spectrum. While the absorption band is rather broad compared to the monomer with its maximum at 738 nm, the emission band is pretty narrow just as the band of the monomeric squaraine with the maximum at 751 nm. This behaviour also supports that exciton coupling causes the broadening of the absorption band of the polymer.[43] Emission occurs only from the lowest energy excitonic state according to Kasha?s rule. However, absorption takes place into several excitonic levels. The fluorescence quantum yield $\Phi_{fl}$ of the polymer is 0.29. 

Upon spincoating a thin film of {\bf 5} on a quartz or ITO (not shown) substrate, the maximum of the absorption is shifted to 757 nm. Interestingly, the shape of the absorption band changed considerably. The small peak at 621 nm which can be seen in the absorption spectrum of {\bf 5} in solution is no longer observable in the absorption spectrum of the film. Furthermore, we investigated the absorption of a film of {\bf 5} on an ITO substrate after annealing at 130\degree{}C for 10 min. These tempering conditions were the same as for the preparation of the solar cell devices which is described below. After annealing the absorption maximum is shifted even further to 778 nm. These effects might be due to a different alignment of the monomer units in a film. According to the structural model explained above, this could result from a more uniform J-type structure with less zig-zag arrangements in the polymer strand. Furthermore, interchain excitonic effects might also contribute to the overall absorption spectrum. In fact, it is known that squaraines can form different kinds of aggregates. Various conditions like substituents and solvents effect the formation of aggregates as well as whether the squaraine is in solution, in a Langmuir-Blodgett film or a microcrystal.[49]

The spectra of {\bf 5} may be compared to spectra in the previous work by Grahn et al. where two-photon absorption of monomeric and thiophene bridged oligomeric squaraines were investigated.[43] In that work, linear optical measurements revealed absorption maxima from 635 nm for the monomer and up to 725 nm for a thiophene-bridged squaraine pentamer with emission Stokes shifts of 240-440 cm$^{-1}$. Extrapolation predicted an absorption maximum of the polymer of 740 nm, close to that of {\bf 5} (738 nm). While the quantum efficiency of {\bf 4} is well in the range of those reported for the thiophene-bridged squaraines by Grahn et al. ($\Phi_{fl}$ = 0.42-0.67), the quantum efficiency of {\bf 5} is less. The squaraines of Grahn et al. were measured vs. Rhodamin 700 as a reference which proved unstable under our experimental conditions though.

\subsection{Cyclic voltammetry}

\begin{figure*}
	\subfigure{\includegraphics[width=8.0cm]{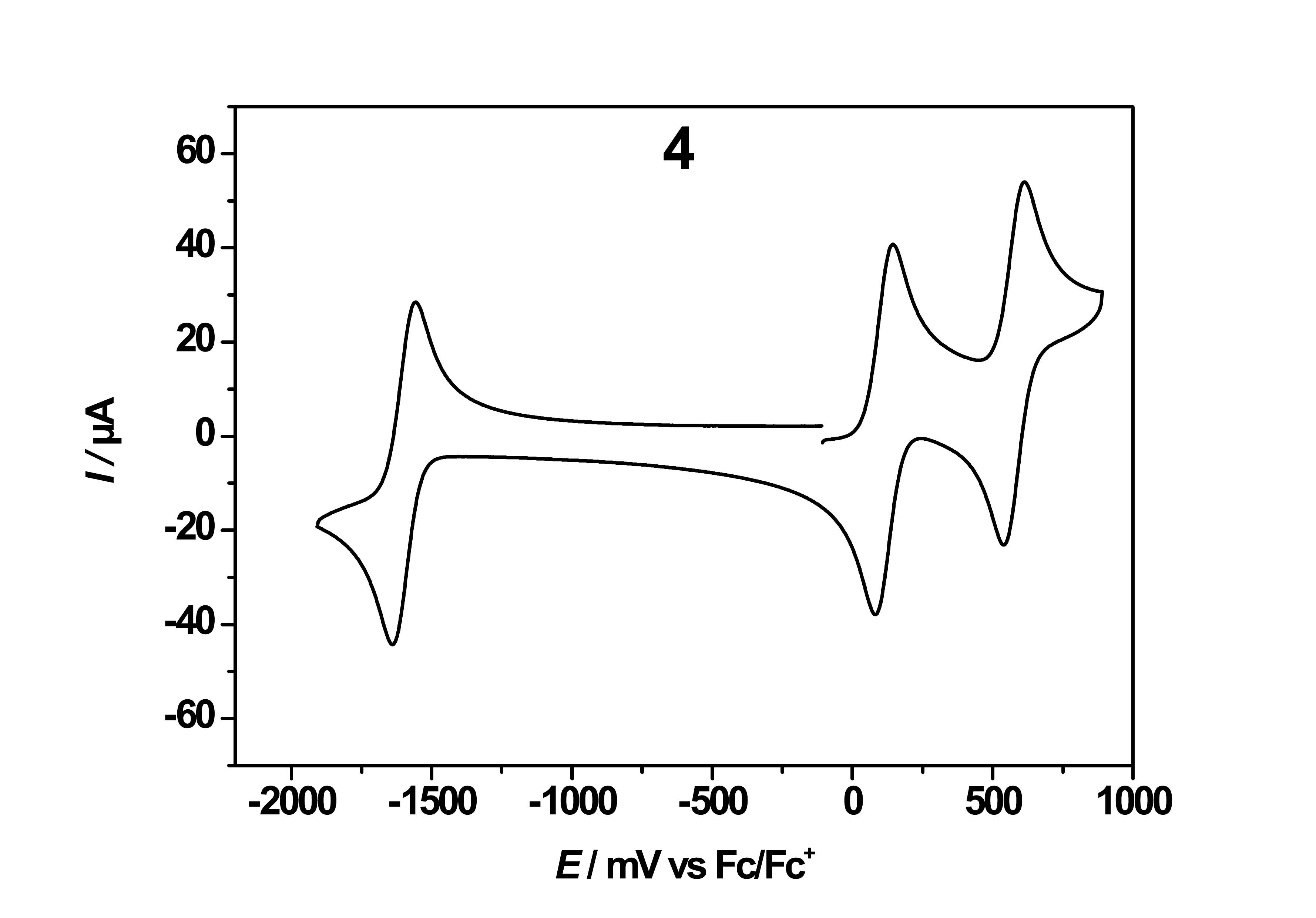}}\quad
	\subfigure{\includegraphics[width=8.0cm]{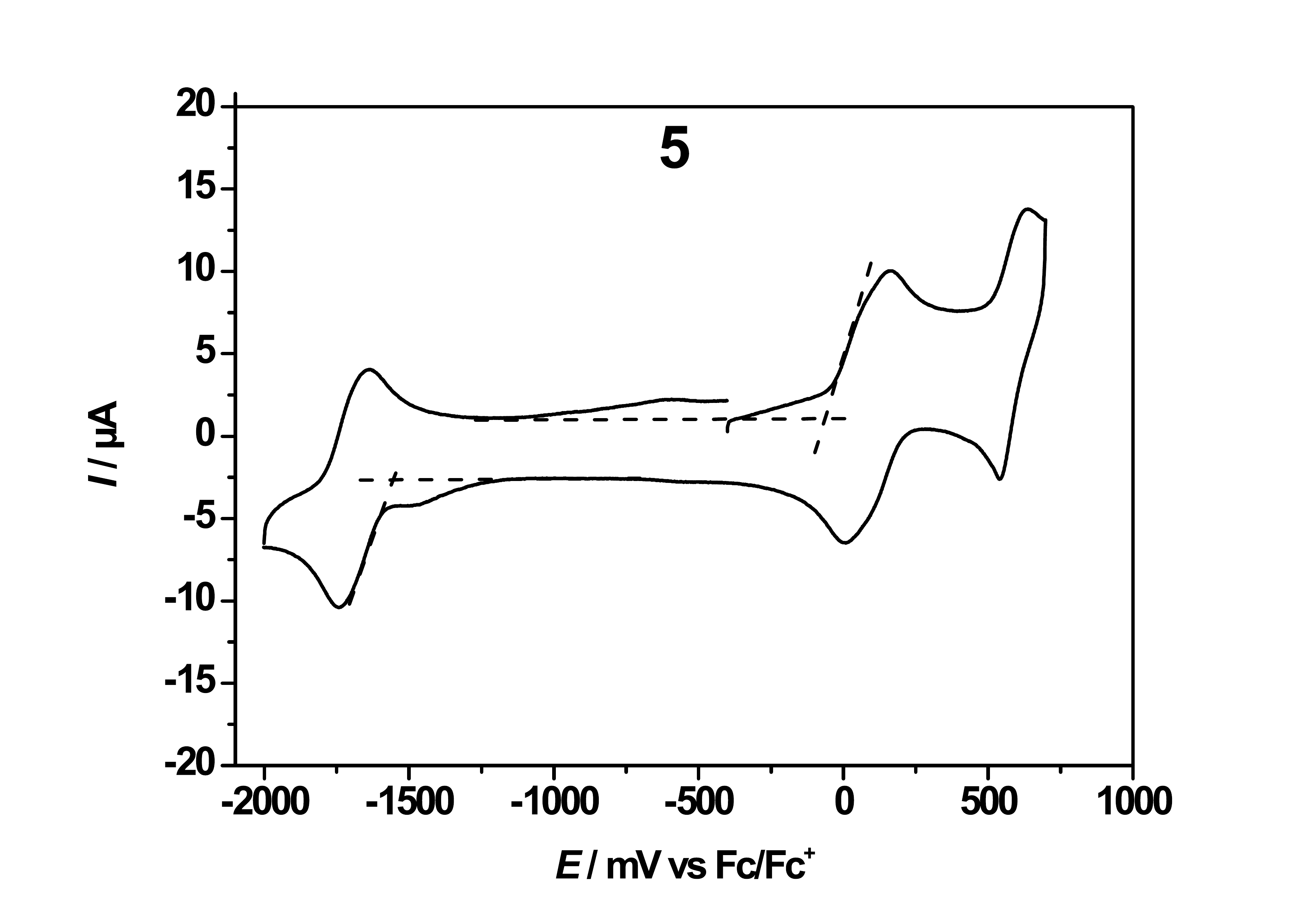}}
	\caption{Cyclic voltammogram of {\bf 4} (left) and {\bf 5} (right) in dichloromethane/0.1 M TBAH at a scan rate of 250 mV s$^{-1}$. %
	\label{fig:3}}
\end{figure*}

We performed cyclic voltammetry (CV) measurements for both the monomer {\bf 4} and the polymer {\bf 5} (Figure~\ref{fig:3}) at room temperature. The potentials were referenced against ferrocene (Fc/Fc$^+$) and dichloromethane was used as solvent with tetrabutylammonium hexafluorophosphate (TBAH) as the supporting electrolyte. The cyclic voltammogram of {\bf 4} shows a reduction at $E_{1/2}$(red) = -1610 mV. Furthermore, there are two oxidation processes at $E_{1/2}$(ox1) = 110 mV and $E_{1/2}$(ox2) = 580 mV. Thin layer measurements revealed that, while the oxidations processes are reversible, the reduction is electrochemically irreversible.

The cyclic voltammogram of {\bf 5} is very similar to the one of {\bf 4}. The half-wave potentials of the reduction is at $E_{1/2}$(red) = -1690 mV, those of the oxidations at $E_{1/2}$(ox1) = 90 mV and $E_{1/2}$(ox2) = 590 mV. However, a closer inspection reveals a distinct broadening of the reduction signal and even a shoulder in the first oxidation wave which we interpret being due to interactions of two adjacent monomer units upon oxidation. Thin layer measurements revealed that reduction and the second oxidation are chemically irreversible. Thus, while there is some interaction detectable in the CV of the polymer, the monomer units still behave as almost independent redox moieties in the polymer.

Analogous to literature procedures we determined the HOMO and LUMO levels for the polymer in solution as well as in the film. We used the onset of the first oxidation and of the reduction to calculate those levels, considering the oxidation potential E = 400 mV of ferrocene versus the normal hydrogen electrode (NHE). The onset potentials found were corrected to NHE and calculated relative to vacuum level by using E(NHE) vs. vacuum = -4.75 eV. This procedure yields HOMO and LUMO levels at -5.09 eV and -3.60 eV, respectively, which results in a band gap of 1.49 eV. Since a straightforward correlation between redox potentials in solution and in the solid state (film) is not obvious, we also investigated the cyclic voltammetry of {\bf 5} as a film on an ITO substrate. The polymer was drop cast on an ITO substrate and afterwards annealed at 130\degree{}C for 10 min analogous to the preparation of the solar cell devices. Here we found HOMO and LUMO levels of -5.14 eV and -3.57 eV, which gives a somewhat higher (by 0.08 eV) band gap (1.57 eV) than in solution. Much in contrast, while generally in good agreement with the electrochemically determined band gap, the optical band gap as determined by the onset of absorption is reduced by 0.08 eV in the annealed film (1.54 eV) compared to solution (1.62 eV).

\subsection{Solar Cells}

Exciton coupling leads to a significant broadening of the absorption bands in the polymer {\bf 5} which prompted us to use this polymer for light harvesting in the low energy region in organic solar cells. Thus, a series of bulk heterojunction devices were built and analyzed at simulated AM 1.5 conditions in this work. The active layer consisted of the polysquaraine {\bf 5} as the electron donor component and PCBM as the acceptor material. For comparison, solar cells with a P3HT/PCBM system were prepared and measured under identical conditions.

The influence of different weight ratios (from 1:3 up to 1:1) of {\bf 5}:PCBM on the solar cell performance was investigated by measuring the current-voltage characteristics. It turned out that blends containing more than 75~\% of PCBM do not show diode characteristics, but are shunted. A reduction of the PCBM fraction leads to an increasing performance of the solar cells. The open circuit voltage $V_{oc}$ does not change significantly, staying at about 475-515 mV (see Table~\ref{tab:1}) for almost all samples. This value is in good agreement with $V_{oc}$ = 540 mV of a polymer/PCBM system calculated by eq.~\ref{eqn:4} 
\begin{equation}
	V_{oc} = \frac{1}{e}	\left( \left| E^\text{Donor}\text{HOMO}\right| - \left| E^\text{PCBM}\text{LUMO}\right|\right) -0.3 \text{eV}
	\label{eqn:4}
\end{equation}
where $e$ is the elementary charge, the 0.3 V represent an empirical factor and -4.3 eV are used for the PCBM LUMO level.[50] Much in contrast the short circuit current density $J_{sc}$ increases considerably from less than 1 mA/cm$^2$ to almost 3 mA/cm$^2$. In combination with a slightly increasing fill factor (up to 34~\%) this results in solar cell efficiencies up to 0.45~\% (see Table~\ref{tab:1}). 

\begin{table}[htdp]
   \caption{Characteristics of {\bf 5}:PCBM solar cells with different weight ratios and a P3HT:PCBM cell as reference at simulated AM 1.5 conditions.}
       \begin{center}
           \begin{tabular}{|c|c|c|c|c|c|}
           	\hline 
               sample	& weight ratio	& $V_{oc}$ / mV	& $J_{sc}$ / mA/cm$^2$ & $FF$ / \%	& $\eta$ / \% \\ \hline \hline
               \multicolumn{6}{|c|}{P3HT:PCBM in chlorobenzene (30 mg ml$^{-1}$)}\\
               1 &1:1 & 600 & 	10.31	& 45.76 & 2.83	\\ \hline
               \multicolumn{6}{|c|}{{\bf 5}:PCBM in chloroform (10 mg ml$^{-1}$)} \\
               2 &1:3 & 500 & 	0.87	& 28.42 & 0.12	\\
               3 &1:2 & 500 & 	2.45	& 31.62 & 0.39	\\
               4 &1:1 & 475 & 	2.82	& 33.92 & 0.45	\\ \hline
               \multicolumn{6}{|c|}{{\bf 5}:PCBM in chloroform (15 mg ml$^{-1}$)}\\
               5 &1:1 & 515 & 	1.36	& 33.49 & 0.23	\\ \hline
           \end{tabular}
       \end{center}
   \label{tab:1}
\end{table}%

\begin{figure}
	\includegraphics[width=8.0cm]{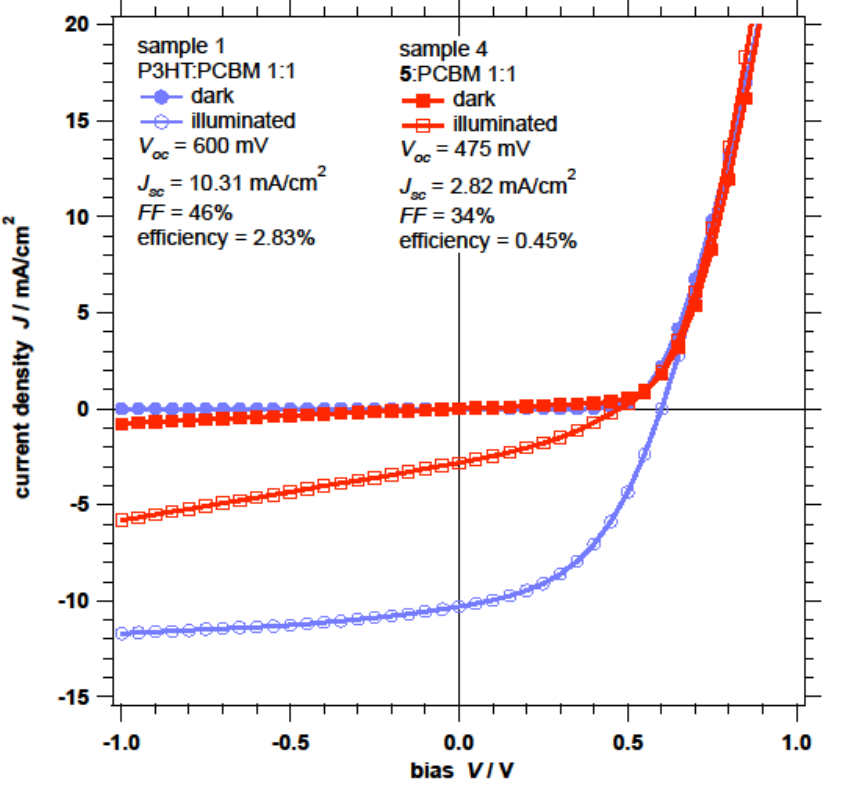}%
	\caption{Current-voltage characteristics of a {\bf 5}:PCBM sample (red lines) and a P3HT:PCBM sample (violet lines) for comparison, both with a 1:1 weight ratio.%
	\label{fig:4}}
\end{figure}

Nevertheless the achieved device performances stay below those of the established and well-optimized P3HT:PCBM system, which is also shown in Table~\ref{tab:1}. Figure~\ref{fig:4} shows the current-voltage characteristics of a 1:1 {\bf 5}:PCBM sample (10 mg m$^{-1}$) in comparison to an identically processed P3HT:PCBM sample (30 mg ml$^{-1}$). 

\begin{figure}
	\includegraphics[width=8.0cm]{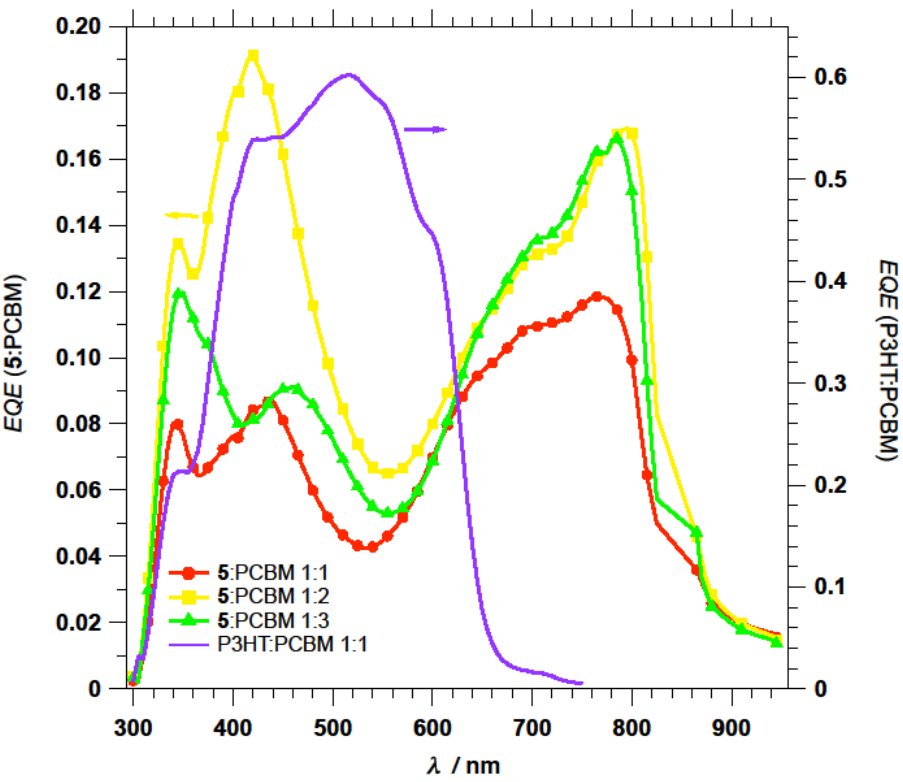}%
	\caption{EQE of {\bf 5}:PCBM samples with varying weight ratios from 1:1 to 1:3 (left axis) and for comparison a P3HT:PCBM sample (right axis)%
	\label{fig:5}}
\end{figure}

The wavelength dependent EQE of {\bf 5}:PCBM solar cells with different weight ratios are shown in Figure~\ref{fig:5} (left axis). For comparison, the spectrum of a P3HT:PCBM reference sample is included (right axis). Compared to the P3HT:PCBM reference cell the squaraine polymer leads to an additional peak between about 600 nm to 850 nm due to its absorption characteristics. This spectral behaviour compares very well with that of the tempered film on ITO (see Figure~\ref{fig:1}). The first peak lies between 300 nm and 500 nm, the EQE therefore showing a minimum at about 550 nm. As expected from the current-voltage characteristics, the EQEs of the {\bf 5}:PCBM cells are below 20~\%, whereas the P3HT:PCBM reference cell reaches over 60~\%.

If we take a close look at the absorption spectra (Figure~\ref{fig:1}) we see how the polysquaraine material becomes eligible for the use in a photovoltaic device in order to harvest light in the near-IR region. The monomer already absorbs intensely in the visible region with the maximum at 645 nm. However, the absorption band is too narrow for being useful as a light harvesting substance since only a small range of wavelengths is covered. The absorption maximum of the polymer is broadened by exciton coupling and further extends into the near-IR region down to 738 nm and even 757 nm when applied as a film on quartz or ITO substrate. After annealing an ITO substrate with the polymer spin coated on top, the maximum of absorption shifted even further to 778 nm with overall contributions from 600 to 850 nm which is in accordance with the EQEs (Figure~\ref{fig:5}). This further broadening is likely due to excitonic interchain interactions in the solid material, typical of squaraine dyes.[23]

\subsection{AFM}

AFM measurements of five solar cell devices have been performed to get a better insight into the morphology of the substrates depending on the weight ratio of the fullerene derivative PCBM. We investigated the morphology of the reference cell P3HT/PCBM with a 1:1 ratio (sample 1) and the {\bf 5}/PCBM cells with ratios of 1:3, 1:2 and 1:1 (sample 2, 3 and 4 respectively) that have been spin cast using solutions with 10 mg ml$^{-1}$. Furthermore we performed AFM measurements of a {\bf 5}/PCBM cell with a ratio of 1:1 that was prepared from a solution of 15 mg ml$^{-1}$ (sample 5).

\begin{figure*}
	\subfigure[Height images of sample 1 (P3HT/PCBM 1:1), roughness: 6.4 nm, RMS: 0.77 nm; average of 4 different images: roughness: 7.9 nm, RMS: 0.80 nm, using 2 $\mu$m × 2 $\mu$m]{\includegraphics[width=12.0cm]{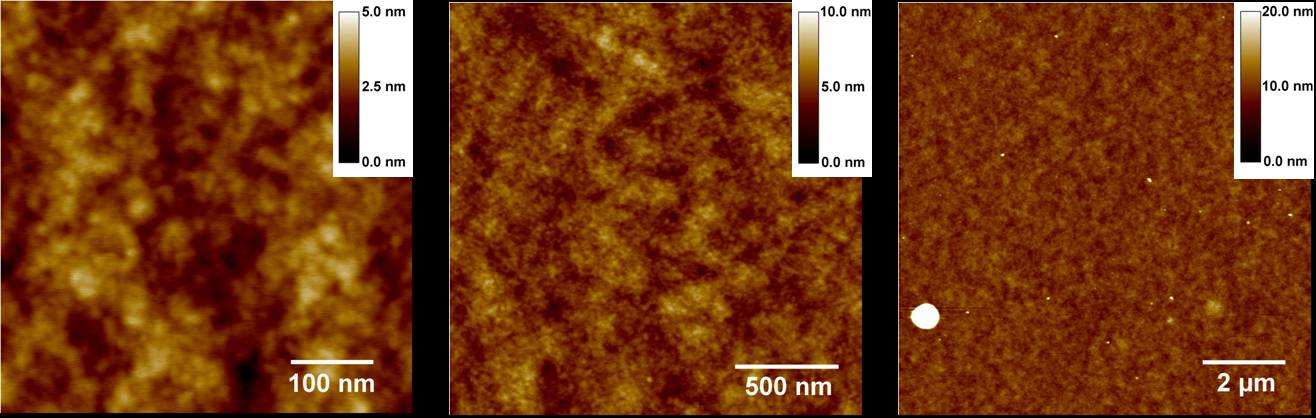}}\\%
	\subfigure[Height images of sample 2 ({\bf 5}/PCBM 1:3), roughness: 16.7 nm, RMS: 1.0 nm; average of 4 different images: roughness: 15.4 nm, RMS: 0.99 nm, using 2 $\mu$m × 2 $\mu$m]{\includegraphics[width=12.0cm]{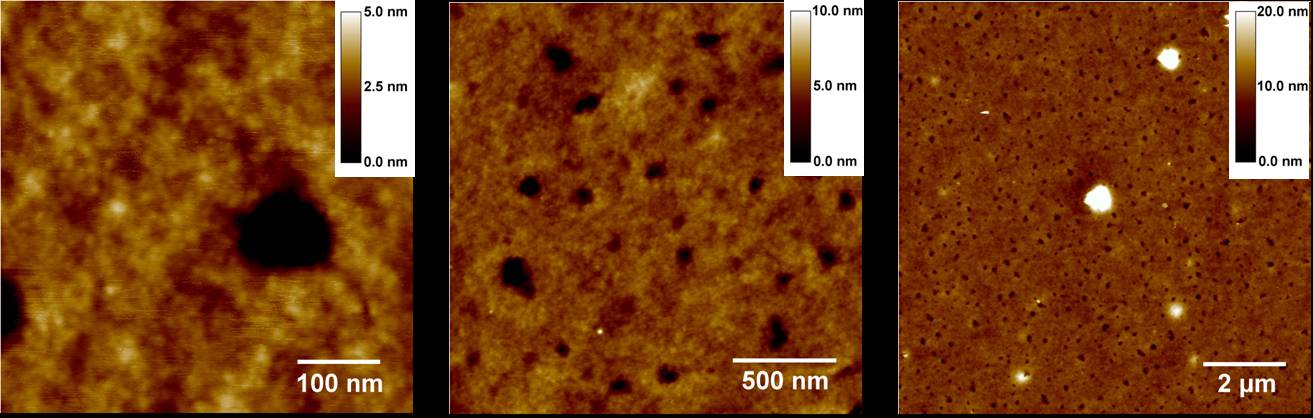}}\\%
	\subfigure[Height images of sample 3 ({\bf 5}/PCBM 1:2), roughness: 6.0 nm, RMS: 0.53 nm; average of 3 different images: roughness: 6.4 nm, RMS: 0.54 nm, using 2 $\mu$m × 2 $\mu$m]{\includegraphics[width=12.0cm]{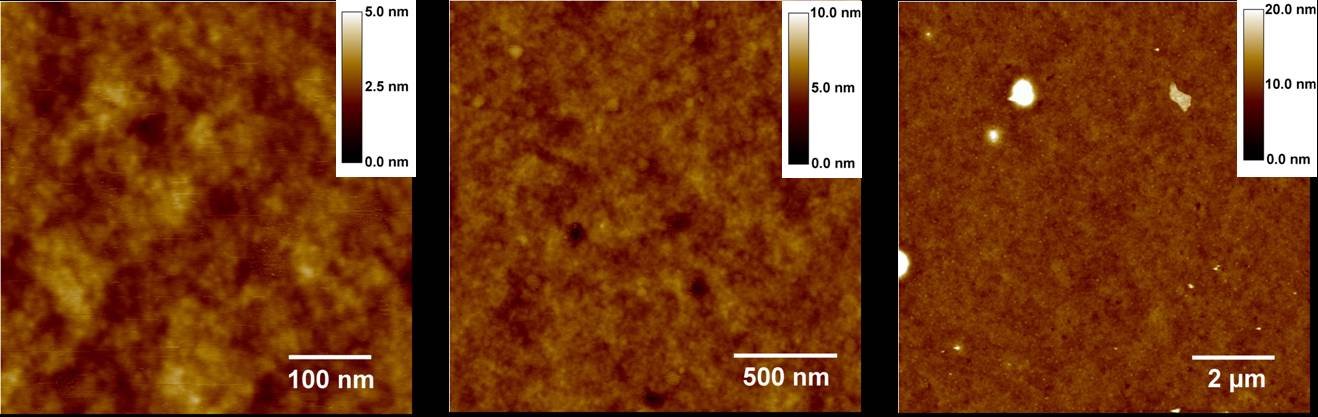}}\\%
	\subfigure[Height images of sample 4 ({\bf 5}/PCBM 1:1), roughness: 8.3 nm, RMS: 0.55 nm; average of 3 different images: roughness: 7.9 nm, RMS: 0.55 nm, using 2 $\mu$m × 2 $\mu$m]{\includegraphics[width=12.0cm]{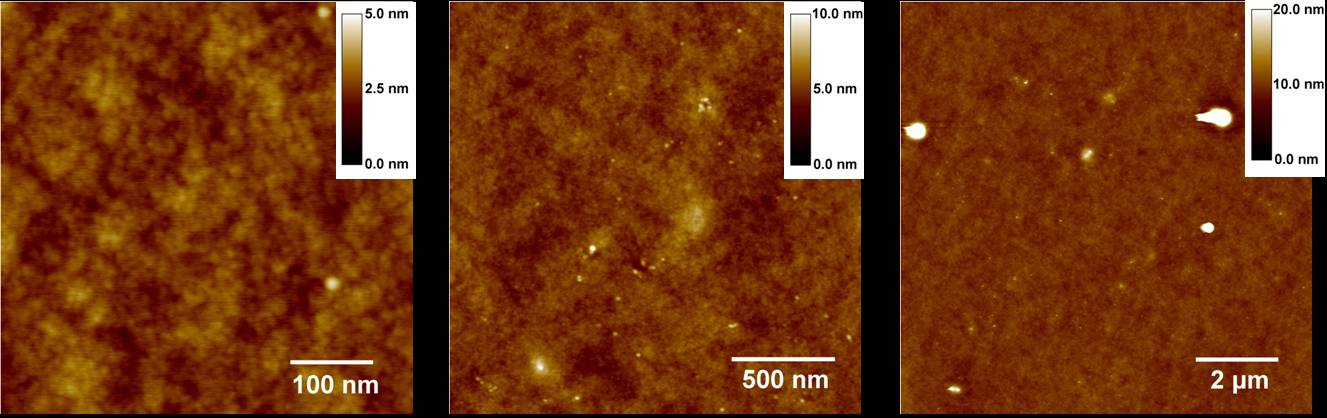}}\\%
	\subfigure[Height images of sample 5 ({\bf 5}/PCBM 1:1), roughness: 6.3 nm, RMS: 0.55 nm; average of 6 different images: roughness: 10.8 nm, RMS: 0.64 nm, using 2 $\mu$m × 2 $\mu$m]{\includegraphics[width=12.0cm]{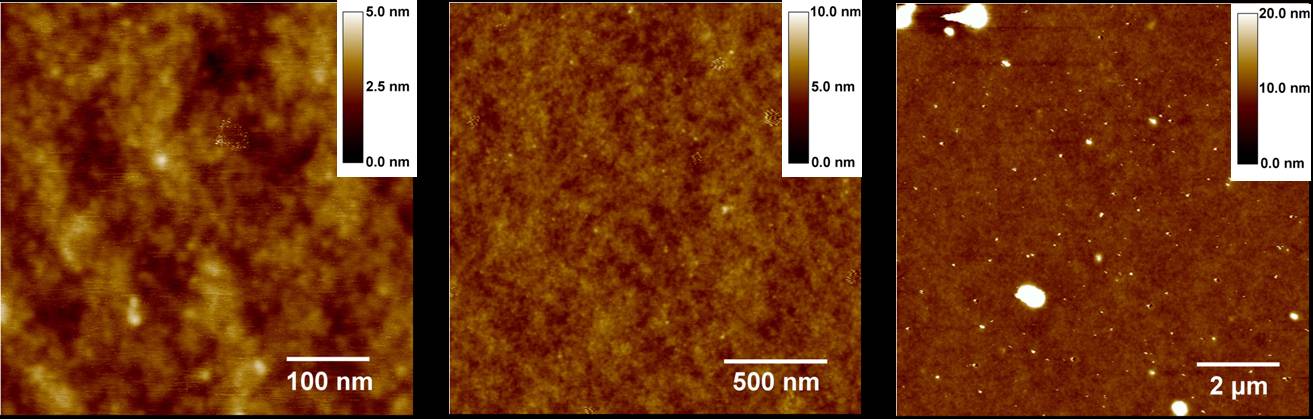}}\\%
	\caption{AFM hight images of solar cell samples 1-5 as given in Table~\ref{tab:1}.%
	\label{fig:6}}
\end{figure*}

The height images shown in Figure~\ref{fig:6}b-e are very similar to the image of the reference cell in Figure~\ref{fig:6}a. All show a rather uniform surface. Some white spots indicate agglomeration which might be responsible for short circuits and the generally weak device performance. Only the images of sample 2 differ considerably compared to the other ones concerning the roughness. In fact there are regular holes of ca 100 nm diameter whose origin is unknown but which might be responsible for the weaker performance of this cell. In general, the morphologies of all samples match with those previously reported for polymer/PCBM blends even though the island structures are not quite as clear as for the reported systems.[51-53] Unfortunately, it is not possible to draw explicit conclusions concerning the relationship of morphology and performance of our devices.

\section{Conclusions}

In this work we built the first bulk heterojunction solar cells based on a system of polysquaraine/PCBM. The EQE measurements proved the wide absorption range of the polysquaraine from 300 nm to 850 nm. This broadened and red-shifted absorption is due to exciton coupling of the squaraine monomers in the polymer strand and due to interchain interactions in the solid material. The non-optimized polysquaraine/PCBM systems achieved power conversion efficiencies of 0.45~\%, which show the potential of the polysquaraine as low band gap absorber for organic photovoltaics. Considering the fact that this was the first polysquaraine studied for the use in bulk heterojunction devices, these are promising results. So far, no structural changes to the polymer have been undertaken to enhance the performance which leaves the chance for further improvement. Structural changes might result in different electronic properties or better solubility which would give us the chance to study the impact of different solvents for the processing. Furthermore, as far as the processing is concerned there is room for improvement since no reference data for such a system exists.

\section*{Acknowledgments}

S.F.V., S.U. and C.L. thank the Deutsche Forschungsgemeinschaft for financing this project within the Graduiertenkolleg GRK 1221. M.L., M.M., C.D. and V.D. thank the Bundesministerium f{\"u}r Bildung und Forschung for financial support in the framework of the MOPS Project (contract no. 13N9867). We thank Prof. F. W{\"u}rthner/ W{\"u}rzburg for the possibility to use his AFM set-up.

\appendix

\section*{Experimental}

\paragraph*{General.} All synthetic preparations were done in standard glassware. The chemicals were obtained from commercial suppliers and used without further purification. Reactions under nitrogen atmosphere have been performed in flame dried Schlenk glassware. Nitrogen was dried over Sicapent from Merck and oxygen was removed by copper catalyst R3-11 from BASF. The solvents were dried applying standard methods and stored under nitrogen atmosphere. Merck silica gel 32-63 $\mu$m was used for flash-chromatography.

\paragraph*{NMR spectroscopy} was carried out either on a Bruker Avance 400 FT-Spectrometer or on a Bruker Avance DMX 600 FT-Spectrometer. Chemical shifts are given in ppm ($\delta$-scale) and adjusted to tetramethylsilane (TMS) as internal standard. Coupling constants are given in Hertz $[$Hz$]$. Abbreviations for spin multiplicities: s = singlet, d = doublet, t = triplet, q = quartet, m = multiplet, dd = doublet of doublet.

\paragraph*{Mass spectrometry} was carried out with solutions of 1 mg/ml in dichloromethane or chloroform on a Bruker Daltonik micrOTOF focus (ESI) or Bruker Daltonik autoflex II (MALDI-TOF).

\paragraph*{UV/vis/NIR spectroscopy} was performed at a Jasco V-570 UV/vis/NIR spectrophotometer. Substances were dissolved in Uvasol solvents from Merck and in case of low solubility filtered over a fibre glass filter GF 92 from Schleicher \& Schuell (Whatman group). 1 cm quartz cuvettes from Hellma were used and the pure solvent was used as reference.

\paragraph*{Emission spectroscopy} was carried out at a PTI (Photon Technology International) fluorescence spectrometer QM-2000-4 including a cooled photomultiplier (type R928P) and a 75 W xenon lamp. Spectra were measured in Uvasol solvents from Merck using 1 cm quartz cuvettes from Hellma. The samples were purged with argon gas for five minutes to remove dissolved oxygen. Oxazine 1 in ethanol was used as fluorescence standard ($\Phi_{fl}$ = 0.11) to define the quantum efficiency.[54] The following equation was used to determine the quantum efficiency.
\begin{equation}
	\Phi_f = \Phi_{f,\text{Ref}} \left( \frac{I(\tilde{\nu}) \times OD_\text{Ref} \times (n_D^{20})^2}{I(\tilde{\nu})_\text{Ref} \times OD \times (n_{D,\text{Ref}}^{20})^2} \right)
\end{equation}
$I(\tilde{\nu})$: integrated emission band, $OD$: optical density of the sample at the excitation wavelength $\lambda_{exc}$, $n_D^{20}$: refraction index. The optical density was determined by the intensity of the absorption band at the excitation wavelength.

\paragraph*{Cyclic voltammetry} was performed with an electrochemical workstation BAS CV-50 W including software version 2.0. We used a three electrodes measuring cell consisting of a platinum working electrode (\o = 3 mm), an Ag/AgCl reference electrode and a helical platinum counter electrode in a cylindrical glass vessel with a Teflon top piece. All measurements were carried out in dried and oxygen free argon atmosphere. Argon was bubbled through the solution for five minutes before the measurement. Tetrabutylammonium hexafluorophosphate (c = $\sim$ 0.1 M) was used as supporting electrolyte and prepared according to known literature procedures. As an internal reference we used ferrocene/ferrocenium. Purification and drying of dichloromethane and acetonitrile was done according to standard procedures. The scan rate was 250 mV s$^{-1}$. For thin layer measurements (20 $\pm$ 5 $\mu$m), the working electrode was put on top of the polished flat surface of a moveable glass hemisphere (\o 8 mm). For the measurements in acetonitrile on the ITO substrate (355.6 mm $\times$ 355.6 mm), which served as the working electrode, the substance was drop cast and tempered at 130\degree{}C for 10 minutes. The conductivity of ITO was 13 $\Omega$/square with a thickness of 125 nm. 

\paragraph*{Gel permeations chromatography (GPC)} was performed with a Jasco PU-2080 Plus and Jasco UV-2077 Plus with polystyrene as standard. The GPC columns (1st: PSS SDV 100 \AA, 2nd: PSS SDV 1000 \AA) were connected to the Jasco HPLC apparatus. The columns were flushed with chloroform for 2-3 h with a flow rate of 0.5 ml/min to 1.0 ml/min until there was a constant baseline signal. Measurements were performed in pure chloroform with a flow rate of 1.0 ml/min.

\paragraph*{AFM measurements} were performed on a Nanoscope IV (Veeco Instruments, Santa Barbara, CA). For AFM investigation an EV-scanner and commercial Si cantilevers with a resonance frequency $\sim$ 330 kHz (Olympus) were used.

\paragraph*{Solar cell preparation and measurement.} For the fabrication of the solar cells we used conductive indium tin oxide/glass substrates. These were cleaned and coated with a thin layer of poly(3,4-ethylenedioxythiophene):polystyrenesulfonate under ambient conditions. Subsequent processing and characterization were performed in a nitrogen atmosphere glovebox. The samples were annealed for 10 minutes at 130\degree{}C before the active layer was spin-coated on top. Both materials, the squaraine derivative as well as PCBM, were dissolved in chloroform. We varied the concentration of the solution (10 mg ml$^{-1}$ and 15 mg ml$^{-1}$) and the weight ratio of the polysquaraine:PCBM blend from 1:5 to 1:1. The coating was followed by another annealing step for 10 minutes at 130\degree{}C and thermal evaporation of the metal contacts (Ca/Al: about 3nm/120nm).

For characterization of the solar cells a Xe-lamp was adjusted to standard testing conditions via the external quantum efficiency (EQE: number of extracted electrons divided by the number of incident photons). This was done by measuring the spectrally resolved photocurrent of the solar cells in comparison to a calibrated photodiode. After calculating the total photocurrent of this cell, the light intensity was adjusted to reach this value with a calibrated silicon reference diode and therefore simulate an illumination which is in good agreement with the AM 1.5 spectrum.[55] For comparison, we also processed and characterized poly(3-hexylthiophene):PCBM samples (both dissolved in chlorobenzene) with the same parameters. 

\section*{Synthesis}

\paragraph*{Synthesis of 5-bromo-1-hexadecyl-2,3,3-trimethyl-3H-indolium iodide 2:} A solution of 5-bromo-2,3,3-trimethyl-3$H$-indole (1.80 g, 7.57 mmol), 1-iodohexadecane (3.20 g, 9.09 mmol) and nitromethane was refluxed for 15 h. After concentrating the mixture under reduced pressure, diethyl ether (50 ml) was added. Upon cooling to 4\degree{}C for 1 h, a precipitate formed which was filtered off, washed with diethyl ether (75 ml) and dried in high vacuum. The product was obtained as a light brown solid (3.30 g, 5.59 mmol, 74 \% yield). M.p. 143-146 \degree{}C. $^1$H-NMR (600 MHz, CDCl$_3$, $\delta$): 7.72 (dd, $^3$J$_{6,7}$ = 8.6 Hz, $^4$J$_{6,4}$ = 1.8 Hz, 1H, Ar H), 7.69 (d, $^4$J$_{4,6}$ = 1.8 Hz, 1H, Ar H), 7.52 (d, $^3$J$_{7,6}$ = 8.4 Hz, 1H, Ar H), 4.71 (t, $^3$J = 7.7 Hz, 2H, N-CH$_2$), 3.09 (s, broad, 3H, CH$_3$) 1.91 (m, $^3$J = 7.4 Hz, 2H, N-CH$_2$CH$_2$), 1.67 (s, 6H, CH$_3$), 1.44 (qui, $^3$J = 7.5 Hz, 2H, CH$_2$), 1.36 (qui, $^3$J = 7.1 Hz, 2H, CH$_2$), 1.31-1.16 (22H, CH$_2$), 0.87 (t, $^3$J = 7.1 Hz, 3H, CH$_3$). 13C NMR (150 MHz, CDCl$_3$, $\delta$): 195.9, 143.5, 140.1, 132.8, 126.9, 124.6, 116.9, 54.7, 50.2, 31.9, 29.70, 29.68, 29.66, 29.65, 29.62, 29.56, 29.46, 29.4, 29.3, 29.2, 28.0, 26.8, 22.7, 23.2, 16.6, 14.1. HRMS (ESI, m/z): [M - HI]$^{\cdot +}$ calcd for C27H45BrIN, 461.26516; found, 461.26525. $\Delta$  = 0.2 ppm

\paragraph*{Synthesis of 5-bromo-1-hexadecyl-3,3-dimethyl-2-methylene-2,3-dihydroindole 3:} 5-Bromo-1-hexadecyl-2,3,3-trimethyl-3H-indolium iodide (3.84 g, 6.51 mmol) was suspended in NaOH (100 ml) and extracted with diethyl ether (3 x 100 ml). The combined organic phases were dried over Na2SO4 and the solvent was removed in vacuo. The crude product was obtained as a yellowish oil (2.93 g, 6.34 mmol, 97 \% yield) which turned red on air. $^1$H-NMR (400 MHz, CDCl$_3$, $\delta$): 7.19 (dd, $^3$J$_{6,7}$ = 8.3 Hz, $^4$J$_{6,4}$ = 2.0 Hz, 1H, Ar H), 7.13 (d, $^3$J$_{6,5}$= 8.3 Hz, $^4$J$_{4,6}$ = 2.0 Hz, 1H, Ar H), 6.37 (d, $^3$J$_{7,6}$ = 8.3 Hz, 1H, Ar H), 3.88 (d, $^2$J$_8$(E),8(Z) = 2.0 Hz, 1H, H(E)-8), 3.84 (d, $^2$J$_8$(Z),8(E) = 2.0 Hz, 1H, H(Z)-8), 3.43 (t, 2H, N-CH$_2$), 1.60 (qui, $^3$J = 7.2 Hz, 2H, NCH$_2$CH$_2$), 1.37-1.22 (32H, 13 x CH$_2$, H-9), 0.88 (t, $^3$J = 7.1 Hz, 3H, CH$_3$).

\paragraph*{Synthesis of 2,5-bis[(5-bromo-1-hexadecyl-3,3-dimethyl-2,3-dihydroindole-2-ylidene)methyl]-cyclobutendiylium-1,3-diolate 4:} A solution of 3,4-dihydroxy-3-cyclobutene-1,2-dione (209 mg, 1.80 mmol) and 5-bromo-1-hexadecyl-3,3-dimethyl-2-methylene-2,3-dihydroindole (1.66 g, 3.95 mmol) in toluene/n-butanol (1:1) (30 ml) were refluxed over night using a Dean-Stark trap. Upon cooling to room temperature, the solvent was removed in vacuo. The residue was purified by flash chromatography on silica gel (gradient: petrol ether (PE)/ethyl acetate (EA) 3:1 $\rightarrow$ PE/EA 2:1 $\rightarrow$ PE/EA 1:1). A concentrated CH$_2$Cl$_2$ solution was filtered and the product was precipitated by dropping a concentrated CH$_2$Cl$_2$ solution into hexane; the precipitate was filtered off and dried in high vacuum. The product was obtained as a shiny red solid (1.01 g, 1.01 mmol, 54~\% yield). M.p. 83-86 \degree{}C. $^1$H-NMR (400 MHz, CDCl$_3$, $\delta$): 7.45 (d, $^4$J$_{4,6}$ = 1.8 Hz, 2H, Ar H), 7.41 (dd, $^3$J$_{6,5}$= 8.3, $^4$J$_{6,4}$ = 1.8 Hz, 2H, Ar H), 6.83 (d, $^3$J$_{7,6}$ = 8.3 Hz, 2H, Ar H), 5.94 (s, 2H, CH), 3.94 (s, broad, 4H, 2 x N-CH$_2$), 1.78 (s, 16H, H-9, 2 x NCH$_2$CH$_2$), 1.45-1.20 (52H, CH$_2$), 0.88 (t, $^3$J = 6.6 Hz, 6H, CH$_3$). 13C NMR (150 MHz, CDCl$_3$, ?): 182.2, 180.1, 169.4, 144.2, 141.5, 130.7, 125.7, 116.6, 110.7, 87.1, 49.3, 43.9, 31.9, 29.68, 29.66, 29.65, 29.63, 29.61, 29.56, 29.49, 29.44, 29.33, 29.32, 27.01, 27.00, 26.9, 22.6, 14.1. HRMS (ESI, m/z): M+ calcd for C58H86Br2N2O2, 1000.50506; found, 1000.50589. $\Delta$ = 0.8 ppm

\paragraph*{Synthesis of the polymer 5:} A mixture of Ni(1,5-cyclooctadiene)2 (160 mg, 598 mmol), 2,2?-bipyridine (93.4 mg, 598 mmol), 1,5-cyclooctadiene (64.0 mg, 598 mmol), degassed toluene (1.0 ml) and degassed dimethyl formamide (DMF) (2.0 ml) was stirred at rt under an nitrogen atmosphere for 30 min. A solution of 2,5-bis[(5-bromo-1-hexadecyl-3,3-dimethyl-2,3-dihydroindole-2-ylidene)methyl]-cyclobutendiylium-1,3-diolate (4) (250 mg, 249 mmol) in degassed toluene (4.5 ml) and degassed DMF (4.5 ml) was added and the resulting mixture was stirred at rt for 6 d. The reaction mixture was poured into MeOH/HCl (20~\%) (4:1) (500 ml) and stirred for 90 min. The purple precipitate was filtered off and washed consecutively with acetone, hexane and THF using a Soxhlet extractor. Each washing was ran over night. The remaining solid was dissolved in CHCl$_3$ (30 ml) and poured into MeOH/H2O (10:1) (440 ml). The resulting precipitate was filtered off and dried in high vacuum. The product was obtained as purple solid (177 mg, 84~\% yield). $^1$H-NMR (600 MHz, CDCl$_3$, $\delta$): 7.54 (s, 4H), 7.05 (d, $^3$J = 7.2, 2H), 6.01 (s, 2H, CH), 4.01 (s, broad, 4H, 2 x N-CH$_2$), 1.87 (s, 16H, H-9, 2 x NCH$_2$CH$_2$), 1.46 (s, 4H), 1.37 (s, 4H), 1.26 (44H, CH$_2$), 0.87 (t, $^3$J = 6.9 Hz, 6H, CH$_3$). Terminal H-atoms could not be detected, most probably due to long chain length. 13C NMR (150 MHz, CDCl$_3$, $\delta$): 182.4, 179.3, 169.6, 143.2, 141.9, 136.7, 126.7, 120.9, 109.7, 87.0, 49.4, 43.9, 31.9, 29.70, 29.69, 29.68, 29.65 (x2), 29.61, 29.56, 29.51, 29.41, 29.35, 27.2 (x2), 27.1, 22.7, 14.1. One C-atom missing. 


\end{document}